\documentclass[preprint,prX,showkeys]{revtex4}
\usepackage[english]{babel}
\usepackage[running]{lineno}
\usepackage{amsmath}
\usepackage{graphicx}
\usepackage[latin1]{inputenc}
\usepackage{pxfonts}
\usepackage{dcolumn}
\usepackage{bm}
\usepackage{color}
\usepackage{units}

\newcommand{\tmop}[1]{\ensuremath{\operatorname{#1}}}

\begin{document}

\preprint{}

\title{Lorentz microscopy mapping for domain walls structure study \\ in L1$_0$ FePd thin films}

\author{Aur\'elien Masseboeuf}
\email{aurelien.masseboeuf@cemes.fr}
\altaffiliation[New ad. : ]{CEMES-CNRS, BP 94347, 29 rue Jeanne Marvig, 31055 Toulouse Cedex}
\author{Christophe Gatel}
\altaffiliation[New ad. : ]{CEMES-CNRS, BP 94347, 29 rue Jeanne Marvig, 31055 Toulouse Cedex}

\author{Pascale Bayle-Guillemaud}
\affiliation{CEA,INAC-SP2M,Laboratoire d'Etude des Mat\'eriaux et des Microscopies Avanc\'ees \\ 17 rue des Martyrs, 38054 Grenoble Cedex 09, FR}
\author{Alain Marty}
\affiliation{CEA,INAC-SP2M,Laboratoire Nanostructures et Magnétisme}
\author{Jean-Christophe Toussaint}
\affiliation{INPG - CNRS/Institut N\'eel, 25 avenue des Martyrs, BP 166, 38042 Grenoble Cedex 9, FR}

\date{\today}

\begin{abstract}
Thin film alloys with perpendicular anisotropy were studied using Lorentz Transmission Electron Microscopy (LTEM). This work focuses on the configuration of domain walls and demonstrates the suitability and accuracy of LTEM for the magnetic characterisation of perpendicular magnetic anisotropy materials. Thin films of chemically ordered ($\unit{L1_0}$) FePd alloys were investigated by micro-magnetic modelling and LTEM phase retrieval approach. The different components of magnetization described by the modeling were studied on experimental images and confirmed by LTEM contrast simulation. Furthermore, quantitative measurements of magnetic induction inside the domain walls were made by using an original method to separate the electrical and magnetical contributions to the phase information. Irregularities were also observed along the domain walls which could play a major role during the magnetization processes.
\end{abstract}

\keywords{Lorentz Microscopy, Transport of Intensity Equation, Perpendicular Magnetic Anisotropy, FePd}

\maketitle
\section*{Introduction}
Lorentz microscopy is used to reveal the magnetic configuration of magnetic domains in a specimen. This powerful technique offers a sensitivity of a few tens of nm.T (this value means that more or less 10 nm of materials with a saturated magnetic induction of 1 T is needed to be observed by LTEM) coupled with the spatial resolution of electron microscopy (chap. 2 of \cite{DeGraef2001}). The basis of LTEM is the deflection of the electron beam by the Lorentz force during transmission through a magnetic specimen. For this reason, only magnetic induction perpendicular to the electron beam can be detected. For in-plane magnetization, a classical LTEM observation (Fresnel and Foucault contrast) reveals black and white contrast representative of domain wall localisation or domains orientation \cite{Chapman1984}. In the case of thin foils with perpendicular magnetization (\textit{i.e.} magnetization parallel to the electron path), this contrast can be obtained by tilting the sample to cause magnetic induction perpendicular to the beam \cite{kohler2000,Hino2004,Aitchison2001}. However, in the perpendicular magnetical anisotropy (PMA) materials class, the domain walls (namely Bloch walls) exhibit an in-plane magnetization component. The specimen could therefore remain untilted, with the magnetization of domains parallel to the electron path and all deflections occurring in the domain walls. To obtain a LTEM signal localized in the domain walls, they must be larger than the spatial resolution of the Lorentz lens or the lens used as an imaging lens (from less than 1 nm for a dedicated Lorentz lens to few nm for a traditional objective minilens). Moreover, the induction perpendicular to the beam must be strong enough to significantly deflect the electron beam (starting with a magnetic sample with a saturated magnetization of $\unit[\mu_0.M_S = 1]{T}$, a thickness of pure perpendicular-to-the-beam magnetization of more than 10 nanometers is needed to deflect significantly the electrons, about \unit[5]{$\mu rad$} in that example \cite{Chapman1984}). In this paper, we present the study of FePd alloys, which, when chemically ordered, exhibit a strong perpendicular anisotropy leading to an up and down domain configuration \cite{Gehanno1997}. Domains are separated by Bloch walls with an in-plane induction around 10 nm in width \cite{Viret2000}. This is large enough to be detected by Fresnel contrast and analysed by mapping methods, as described in the next section.\\
Fresnel and Foucault contrasts are both standard imaging modes but neither of them gives direct quantitative information about the magnitude of magnetization. Image series are needed to quantify the observed deflections. Differential Phase Contrast is a mapping technique that has been implemented initially in a STEM microscope \cite{Chapman1978} and then extended to conventional TEM \cite{Daykin1995}. This technique is an extension of imaging by Foucault contrast and has not been used in this work. Using a wave-mechanical approach, the action of the magnetic induction on the electron wave is modelled as phase shift. This is the well-known Aharonov-Bohm effect \cite{AHARONOV1959}. Although phase information is lost by image formation, focal series allow phase retrieval by the solving of the Transport of Intensity Equation (TIE) \cite{Teague1983}. The TIE equation,
\begin{equation}
\label{Eq1}
			\frac{2\pi}{\lambda}\frac{\partial}{\partial z}I_{x,y}  = -\nabla_{xy}\cdot \left[  I_{x,y} \nabla_{xy}\phi \right],
\end{equation}
relates the phase shift to image intensity along the optical axis, with $\lambda$ being the wavelength of the electrons, $I_{x,y}$ being the intensity distribution, $\phi$ being the phase shift of the electron wave and $\nabla_{xy}$ is the in-plane gradient \cite{Paganin1998}. Nowadays this technique has been widely developped, by means of numerical calculations \cite{De2001,Volkov2002} or precautions in the use of mathematical approximation \cite{Beleggia2004,McVitie2006,Petersen2007,Koch2008}. The technique has been applied to a wide variety of both magnetic \cite{Volkov2004} and non-magnetic materials \cite{Petersen2008}.\\
The variation of intensity along the z axis is related to the gradient of the phase shift. As it has been described by Aharonov and Bohm, the phase shift is the resuolt of two contributions : an electrical one, due to the mean inner potential of the material and a magnetic one, due to the magnetic induction distribution in and around the sample, namely
\begin{equation}
\label{Eq2}
			\phi\left( x,y\right)  = C_{E} \int V_{int} (x,y) dz - \frac{e}{\hbar}\int_{\infty} A_{z}(x,y,z) dz ,
\end{equation}
where $C_{E}$ is electron energy ($E$) dependent ($ C_E = 2\pi e/ \lambda E \cdot (E + E_0 )/ (2E + E_0)$ with $E_0=m_0c^2$ the rest energy of electrons), $ V_{int} $ is the mean inner potential of the sample (the electrical phase shift origin). $ A_{z} $ is the z-component of the magnetic vector potential describing the magnetic induction distribution in a plane (for a given z) perpendicular to the optical axis ($\vec{B} = \vec{\nabla} \times \vec{A} $).\\
Knowledge of the phase shift gives access to quantitative magnetic information if the electrical contribution can be removed. Various techniques have previously been used to achieve this\cite{Dunin-Borkowski2004}. Most of these methods need the acquisition of two different phase informations (changing the sample orientation, saturation condition or accelerating voltage for example). In this work, a novel, easily implemented method, with only one phase calculation is proposed for thin foil samples prepared with the classical method (mechanical polishing and ion milling).

\section*{Calculation}
Consider the definition of the magnetic vector potential, the z component variation in the $x,y$ directions can be expressed by :
\begin{eqnarray}
\partial_y A_z = B_x + \partial_z A_y \nonumber\\
\partial_x A_z =- B_y - \partial_z A_x .
\end{eqnarray}
Integrating over all the z-axis directions yields,
\begin{eqnarray}
  \int_{\infty} \partial_y A_z (x, y, z) dz = \int_{\infty} B_x (x, y, z) dz -
  \left[ A_x (x, y, z) \right]_{z = - \infty}^{z = \infty} \nonumber\\
  \int_{\infty} \partial_x A_z (x, y, z) dz = \int_{\infty} - B_y (x, y, z) dz
  - \left[ A_y (x, y, z) \right]_{z = - \infty}^{z = \infty}
\end{eqnarray}
Assuming that magnetic vector potential vanishes at the infinity (or is the same on each side, far from the sample), one gets
\begin{equation}
\label{Eqfinale}
  \frac{e}{\hbar} \: \vec{k} \times \vec{b}(x,y)  = - \vec{\nabla} \cdot \phi \left( x, y \right) + C_E \vec{\nabla} \cdot
  \left[ \overline{V (x, y)} t (x, y) \right]
\end{equation}
where $\vec{k}$ is a unit vector of the optical axis. The main effect of the vectorial product on the \emph{l.h.s.} of Equation \ref{Eqfinale} is the inversion the x and y components of $\vec{b}(x,y) = \int_{\infty} \vec{B}_{} (x, y, z) \tmop{dz}$, the integration along the optical axis of the magnetic induction including magnetization in the sample and stray fields. $\overline{V (x, y)}$ is the integration of the mean inner potential over all the thickness of the sample.\\
It is clear that for a sample of constant thickness and composition, the electrical contribution disappears, and for a uniform gradient of thickness and uniform composition, the electrical contribution is reduced to a constant. Hereafter, we show how this equation can be used to remove the electrical contribution from experimental phase measurements obtained using conventionnal electron microscope specimens.

\section*{Experiment}
This paper presents results obtained from a sample with PMA configuration, exhibiting two different in-plane contributions. Previous LTEM observations have been carried out on similar alloys \cite{Aitchison2001} and the authors reported that it was necessary to tilt the sample in order to reveal the domain structure. In our case the in-plane components lead to strong Fresnel contrast without tilting and we were able to reveal the domain wall configuration using reconstruction by TIE solving. The information given by the phase retrieval is thus directly related to domain wall information and not to the domain magnetization, parallel to the electron beam. Such dependance enables a quantitative exploration of the domain walls rather than the whole sample.\\
The sample was grown on a MgO (001) substrate by Molecular Beam Epitaxy (MBE) according to the following sequence : a thin Fe layer (2 ML) was deposited  in order to initiate the epitaxial growth on the substrate, a16 nm chemically disordered FePd$ _{2}$  layer was first codeposited as a magnetic layer having a vanishing anisotropy. Then, a FePd ordered (L1$_{0}$) layer (37 nm) with a strong PMA was added at 350$^{\circ}C$. The film was then covered with a 1.5 nm Pt capping layer to prevent oxidation. The structural and magnetic properties of this type of alloy can be found elsewhere \cite{Gehanno1997}. Sample for electron microscopy was prepared using a classical method by mechanical polishing and Ar ion milling. The preparation was in a plan-view geometry and the angle used for ion milling was 6 degrees. The microscope used for the study was an FEI TITAN equipped with a dedicated Lorentz lens operating at 300 kV and fitted with a Tridiem GATAN Imaging Filter. All images were zero-loss filtered and data processing has been completed by scripts we have developped in Digital Micrograph from Gatan. The TIE operation was carefully done using the fourier approach\cite{De2001}, enhanced by image symetrization\cite{Volkov2002}. Micromagnetic simulations were performed with GL\_FFT (copyright CNRS, Institut Néel Grenoble \cite{Toussaint2002}). The micromagnetic configuration was used to simulate Fresnel contrast and compared to experimental images.

\section*{Results}

\begin{figure}[!h]
\includegraphics[width = \columnwidth]{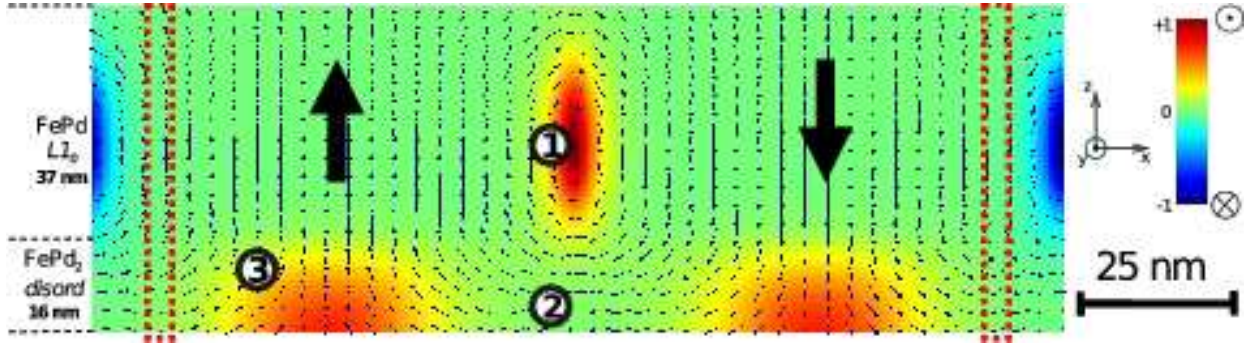}
\caption{\label{Fig1}Micromagnetic simulation from GL\_FFT on the FePd bi-layer. The input parameters are displayed in the text. The simulations contain only the magnetic layers. The growth direction is along the vertical. A colour scale is used for the component of magnetization perpendicular to the figure. Marks on the figure show the in-plane magnetic components described in the text. Dashed areas correspond to some areas which fulfil the no-deflection conditions.}
\end{figure}

The focus here is on micromagnetic simulations carried out for this magnetic stacking (see Fig 1). The simulations are based on the temporal integration of the Landau-Lifshitz-Gilbert equation using regular prisms to discretize the sample. The input parameters used for FePd were chosen as follows \cite{Gehanno1997} : Exchange constant A = $\unit[6.9 \: 10^{-12}]{J.m^{-1}}$, Uniaxial Anisotropy K = $\unit[1.03 \:10^{6}]{J.m^{-3}}$, Saturated Magnetization $\mu_{0}$M$_S$ = $\unit[1.294]{T}$. The cells are $ \unit[0.781]{nm} \times \unit[0.625]{nm}$ and infinite along the \emph{y} direction. The results are presented here in a cross-sectional view, with the growth direction along the z axis.\\
The top magnetic layer with a L1$_{0}$ structure can now be observed. This layer exhibits an out-of-plane magnetic anisotropy which can be described  as a succession of up and down domains separated by Bloch walls (mark 1 on Fig \ref{Fig1}) and with a periodicity of $\unit[160]{nm}$. These domain walls are surrounded by Néel Caps due to the demagnetizing field. 
The bottom magnetic layer is the soft layer, where the magnetization tends to be in-plane. This implies a larger Néel Cap below the domain walls (mark 2). In-plane components of the Néel Caps are usually considered to be symmetric : in terms of electron deflection the two deflections are cancel one another. In this sample the balancing no longer exists. Below the domains (mark 3), the magnetization enhancement shown previously (mark 2) gives rise to another in-plane component. This component is parallel to the domain walls and generates a magnetic deflection. Considering the in-plane induction sensitivity of LTEM, three magnetic components considered to be acting on the electron beam in this sample : (1) Bloch walls, (2) bottom Néel Cap, (3) soft layer under the domains. In our 2D case, the second term deflects the electrons along the y direction. The technique is not sensitive to that magnetization component because the deflection happens along the Bloch wall. It would only be apparent  at the point of a sudden change in the wall direction. \\

\begin{figure}[!h]
\includegraphics[width =  0.7\columnwidth]{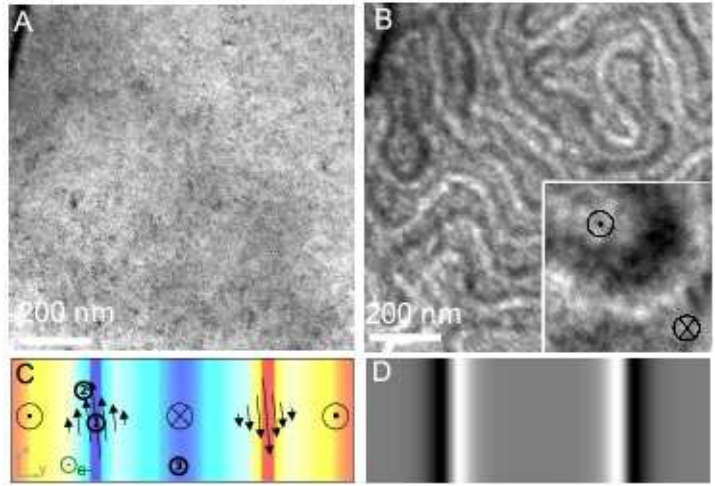}
\caption{\label{Fig2}\textbf{A.} In-focus image of FePd sample. \textbf{B.} Over-focus ($15 \mu m$) of the same area as in A. The inset is a magnified view of the contrast of a domain wall. Arrows pointing the domains were choosen arbitrarily. \textbf{C.} Idealized profile of the perpendicular magnetic anisotropy system in plane view geometry. Domains are separated by Bloch walls. The marks 1-3 correspond to the same magnetic components as in Fig \ref{Fig1}. The same color scale as in Fig \ref{Fig1} is also used. \textbf{B.} Corresponding Fresnel contrast obtained by simulation.}
\end{figure}

Magnetic Fresnel contrast is observed on the defocus image (Fig \ref{Fig2}.B) for an untilted sample. This contrast comes directly from the domain wall area (mark 1 on Fig \ref{Fig2}.C) where the magnetization lies in-plane. No deflection occurs in the domains themselves. The typical Fresnel contrast for a Bloch wall is a couple of bright and dark bands revealing a continuous variation of in-plane magnetization (see Fig \ref{Fig2}.D). As shown by the simulation, contrast coming from the soft layer (mark 3) is hidden by the intensity of the Bloch wall contrast.\\

We performed a phase retrieval process via the solution to the TIE (Fig \ref{Fig3}). The method employed here was the numerical method described by De Graef et al. in \onlinecite{De2001}. Over- and under-focused images used for the reconstruction are presented in Fig \ref{Fig3}A-B. The calculated phase (Fig \ref{Fig3}-C) exhibits two kinds of variation. The first is a continuous gradient along the phase image of about 10 mrad (along the diagonal of Fig \ref{Fig3}.C, from dark to bright contrast). Smaller variations are also observed in the same image, which are measured around 0.1 mrad. The gradient in the vertical and horizontal direction gives a vectorial distribution which can be displayed as a scattergram (inset of Fig \ref{Fig3}.C). The scattergram is a 2D representation of the distribution of the magnetic induction vectors, \textit{i.e.} a bidimensionnal histogram of the magnetic vectors distribution (see also appendix). As deduced from Eq. \ref{Eqfinale} there is both an electrical and a magnetic contribution to this scattergram (see also Fig \ref{Fig3}.E). Stated a constant variation of the thickness profile due to the sample preparation (this has been checked using the inelastic electron ratio \cite{Malis1988} on the sample, results not presented here) the electric contribution is reduced to a constant and the magnetic information is therefore described by
\begin{eqnarray}
   \phi_{Mag} & = & \phi_{Measured} - \phi_{El} \\
   \vec{k} \times \overrightarrow{B_{\bot}} & = &\overrightarrow{V_{calc.}} - \overrightarrow{cste} .
\end{eqnarray}
$\overrightarrow{V_{calc.}}$ describes the vector field deduced from the phase gradients (and presented as a scattergram in the inset of Fig \ref{Fig3}.C). The $\overrightarrow{cste}$ represents the vectorial contribution (derived from the phase) which has no magnetic origin.\\
In the scattergram, the gradient of the magnetic phase leads to magnetic induction vectors (with coordinates that are rotated by 90$^{\circ}$ with the use of $\vec{k} \times$) while the gradient of the electrical phase is the same for each pixel of the phase image. As a result, each pixel of the gradient-phase image can be described as a sum of a constant vector due to the electrical contribution and a vector due to the magnetic contribution. A straightforward way to remove the electrical contribution is to remove the vectorial constant due to the electrical contribution (see the scheme in Fig \ref{Fig3}.E). This is done here by considering an isotropic distribution of magnetic moments in the image, which has been checked on previous MFM measurement \cite{Gehanno1997a}. Up and down domains are interlaced so the domain walls orientations are pointing in all in-plane directions. We note there that a previous knowledge (MFM observation in that case) of the magnetic distribution could be a key point for such a quantitative analysis.\\
The scattergram (inset of Fig \ref{Fig3}.C) exhibits two main characteristics : a round shape due to the distribution of isotropic magnetic moments, and a center of shape which does not correspond to the center of the scattergram image. When re-centering the distribution of magnetic vectors in the scattergram image, the electrical contribution to the phase is removed (Fig \ref{Fig3}.D). The small phase variations due to magnetic deflections are now easy to observe.\\
Another simple way of removing this component was previously proposed by Volkov and Zhu in \cite{Volkov2004}. The aim of this method was to estimate the electrostatic phase by fitting a parameter corresponding to the log-ratio equation \cite{Malis1988}. This method also needed only one phase information and no previous quantitative thickness datas. Nevertheless, it is necessary to add another mathematical approach, namely fitting the parameter for electrostatical phase shift approxiamation. Furthermore this technique is only working with sharp edges in the imaging area. It could not have been implemented here. The two techniques discussed here can thus be described as complementary techniques for magnetic quantitative analysis in LTEM.\\
We now discuss about the MIP variations in the image. The large field of view used in LTEM observation (around 1 $\mu m$ in that case) and the geometry of the TEM sample (prepared with traditional ion milling) leads to smooth variations of the MIP in the image. As a simple evaluation, the MIP here could not be evaluated with a precision of more than 1 V (see for example literature review for MIP of semiconductors in ref. \onlinecite{Kruse2006}). Over the field of view of that observation the MIP is mainly modified by the ratio of MgO over the global MIP of the magnetic layer. Considering the 6 degree angle used for sample preparation, the variation in the MIP value corresponds to less than 1 V (considering MIP$_{MgO}=14~V$ and MIP$_{FePd}=25~V$) which is less than the precision we could expect on such a MIP estimation. Then small variations of the MIP (due to grain boundaries, atomic steps or structural defects) are not considered in our experiment.\\

\begin{figure}[!h]
\begin{center}
\includegraphics[width = 0.6\columnwidth]{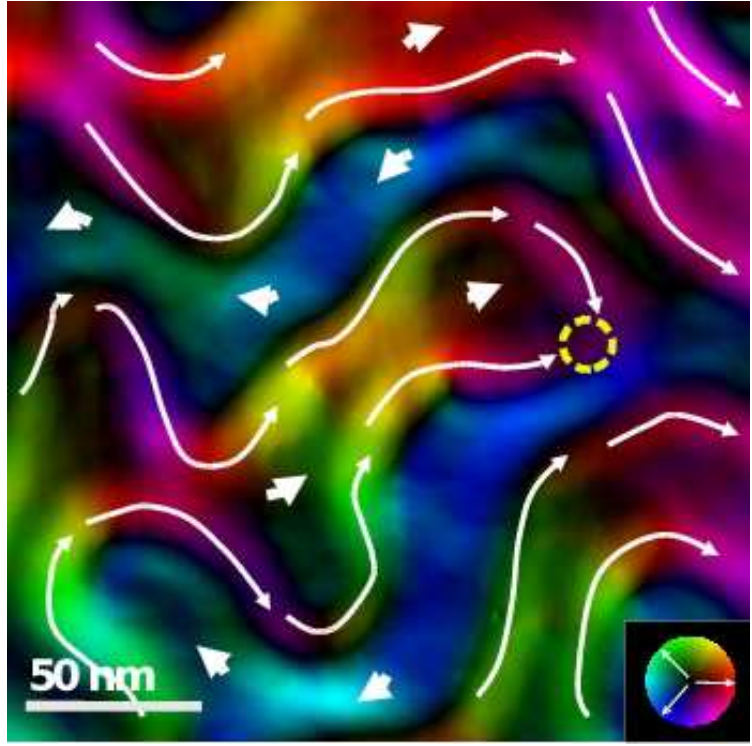}
\caption{\label{Fig4}Magnetization map from a phase retrieval process issued from a zoom in Fig \ref{Fig3}. The color scale used here shows the magnetic vector direction with a colour and the intensity of the colour corresponds to the modulus of this magnetic vector (see scale as inset). Thin and long white arrows are pointing the magnetization in the domain walls. Thick and small white arrows are pointing the magnetization laying under the domains. The dashed circle points a singularity in the wall described in the text.}
\end{center}
\end{figure}

\begin{figure}[!h]
\includegraphics[width = \columnwidth]{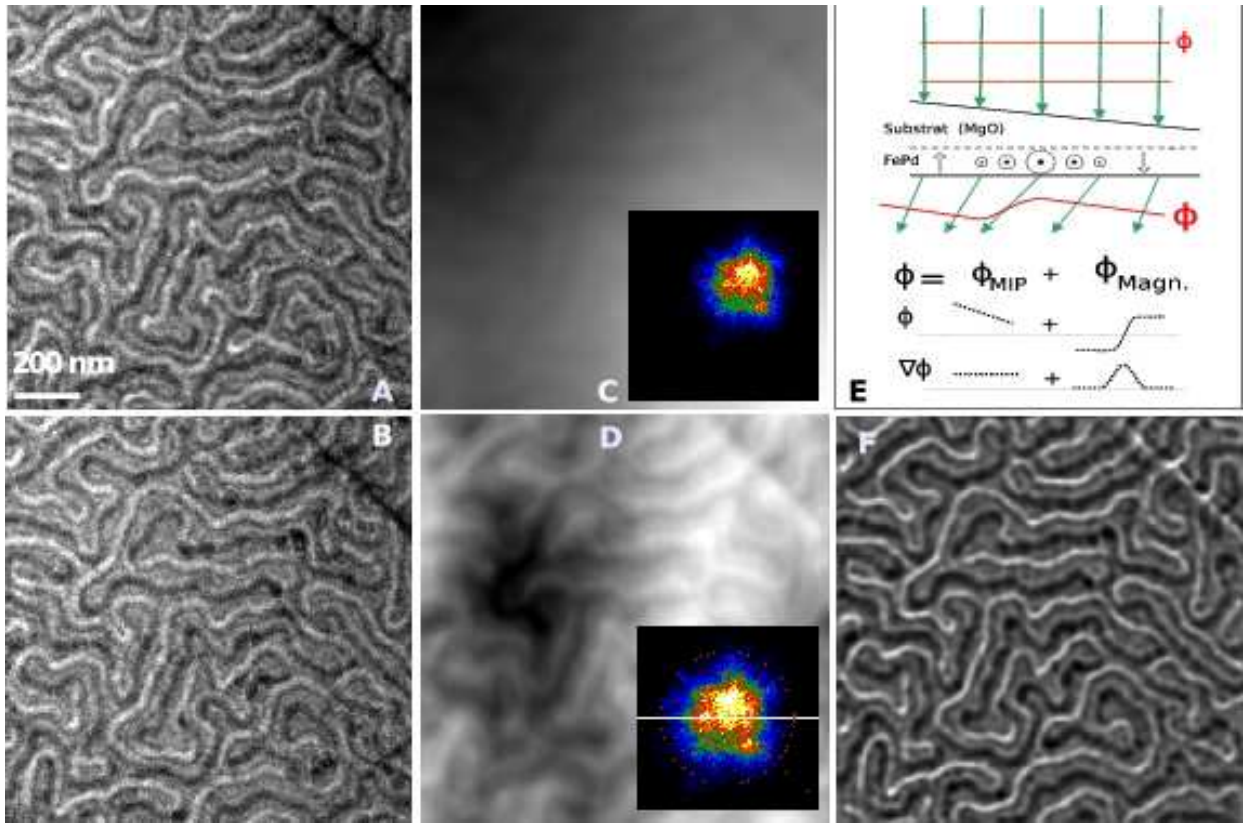}
\caption{\label{Fig3}\textbf{A.} Under-focused ($\unit[-20]{\mu m}$) image used for the phase retrieval process. \textbf{B.} Corresponding over-focused ($\unit[20]{\mu m}$) image. \textbf{C.} Phase image obtained after the reconstruction from focal serie. The inset is the scattergram associated to the phase gradient (in temperature scale). \textbf{D.} Phase image and its corresponding scattergram after electrical contribution removal. \textbf{E.} Explicative scheme of the phase shift origin. The phase of the electron wave (red line) is shifted by an electrical and a magnetic origin. The decomposition of $\phi$ and its gradient (directly related to $\vec{B}$) is explained. \textbf{F.} Fresnel contrast simulation ($\Delta f = \unit[20]{\mu m}$) obtained from the magnetic induction map determined from phase image. The domain walls appear as coupled dark-bright bands.}
\end{figure}

We have then determined the pure magnetic induction map. A zoomed area of the reconstruction is shown in Fig \ref{Fig4}. In that representation we observe two main areas where the magnetization is in-plane :
\begin{itemize}
\item "Magnetic tubes" which are running all around the sample. These are the domain walls (intense colour on the map Fig \ref{Fig4}, enhanced by long thin arrows).
\item Other areas where the magnetization is also in-plane with a modulus which is varyng along the layer (and located around the domain walls) correspond to the domains (weaker colour on the map Fig \ref{Fig4}, enhanced by small, wide arrows). The signal does not come from the out-of-plane domains but in the soft layer where the longitudinal magnetization area is parallel or antiparallel to the domain walls (mark 3 on Fig \ref{Fig1}).
\end{itemize}

We might expect some dark areas corresponding to pure perpendicular magnetization. These areas are not localised in the middle of the domains (due to the soft layer) but on the boundaries between domains and domain walls (see dashed areas in Fig \ref{Fig1}). The no-deflection condition can be explained by a balancing between the Bloch domain wall (mark 1 on Fig \ref{Fig1}) and the longitudinal magnetization area under the domain (mark 3). This balancing occurs only if the domain wall is running in the opposite direction to this longitudinal magnetization area. The dark areas on the color map (Fig \ref{Fig4}) do not always surround domain walls but only when a domain wall and the longitudinal magnetization area under the domain near this wall are antiparallel.\\
If we assume that stray fields and Néel Caps are cancelling one another on both sides of the sample, we can evaluate magnetization inside the domain walls. Thus, considering a constant magnetic thickness of 37 nm at the domain wall, we found a value of magnetic induction around 1.3 T in the domain walls (long thin arrows in Fig \ref{Fig4}). This value is according to the FePd(L1$_0$) saturated magnetization (measured to be 1.29 T in \cite{Gehanno1997a}). On the contrary, considering a magnetic thickness of 16 nm between the domain walls (the magnetic induction in the L1$_0$ layer is purely perpendicular) we measured a value betweenn 0.01 and 1.0 Tesla in the soft layer (small and wide arrows in Fig \ref{Fig4}). These values are in the range expected for the disordered FePd$_2$ saturated magnetization. They could be over-estimated when measured close to the domain walls.\\
Domain wall widths appear to vary, depending on shape (straight or curved) and the proximity of other domain walls. We can assume a size around 10 nm for the domain walls in this sample. Theorical values of 8 nm \cite{Viret2000} were calculated using an infinite problem without considering stray field energy or the shape of the thin foil. Nevertheless, this kind of value has to be evaluated for different values of defocus by means of \emph{zero-defocus} approximation. MFM measurements previously showed a pseudo-period for this sample of 160 nm, which is the same as our measurements with an auto-correlation.\\

Along the walls, "switching" points were observed where the magnetization abruptly turns $180^{\circ}$ with respect to the wall direction. One of these points is indicated in the Fig \ref{Fig4} with a dashed circle. It can also be observed in the Fresnel contrast as an inversion from dark-bright contrast to bright-dark contrast. These singularities could be of great interest if it could be confirmed that they are Vertical Bloch Lines \cite{HUMPHREY1985,Thiaville2001}. The Bloch lines observed in traditional in-plane anisotropy materials play a major role during the magnetization process. This role will be discussed in a future publication on the nature and the behaviour of these lines.

\section*{Conclusion}
We have demonstrated that Lorentz Microscopy is potentially a powerful tool for the investigattion of magnetization configuration at the magnetic domain walls scale. It is possible to study materials which are anisotropic out-of-plane without tilting the specimen. FePd alloys have been studied and their Bloch wall structure revealed. We have noted some defects that appear along the domain walls which are potentially important during the magnetization process. Observations have been carried out using a new method for subtracting mean inner potential information from phase datas. This subtraction enables the quantification of the magnetic map with a good correlation to previous measurements. Resolution of the Transport of Intensity Equation is a very high-resolution technique both spatially (around ten nanometers) and in terms of magnetic sensitivity (few tens of nm.T). 

\section*{Acknowledgments}
We grateful thank Dr. P. Cherns for critics and useful remarks along the redaction of this manuscript.
\newpage
\section*{Appendix}
\begin{figure}[!h]
\includegraphics[width =  0.5\columnwidth]{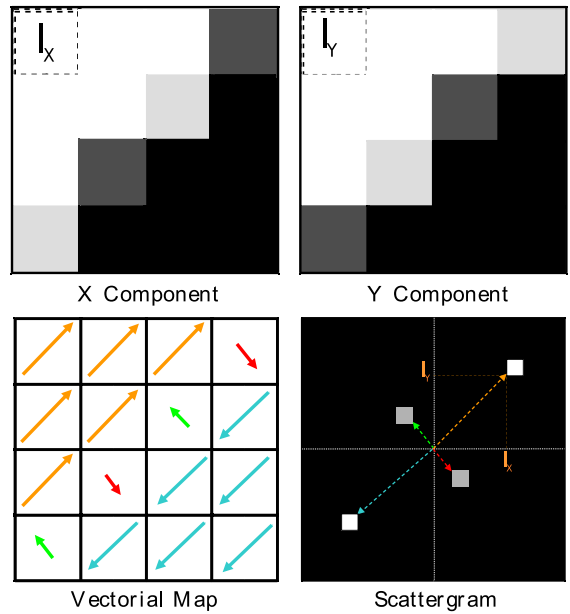}
\caption{\label{Fig5} Scattergram for a 16-pixels image representing a schematic Cross-Tie wall. The two top images are the X and Y components of the magnetization. The bottom left image is a vectorial image representation of the magnetization distribution.The bottom right image is the scattergram. The color scale used in the scattergram is zero for black and the gray levels correspond to the among of times the corresponding vector is dispalyed in the previous images. Colors are used to show the position of the vectors in the scattergram.}
\end{figure}

We describe here the notion of scattergram which can be labeled as a \emph{bidimensionnal histogram of the moment vectors} by gray level scale. As shown in Figure \ref{Fig5}, we use a 2D representation to count the magnetic moments in a vectorial map. Starting with two images, displaying two components of magnetization (see Fig \ref{Fig5} upper left and upper right) a vectorial representation can be made (see Fig \ref{Fig5} bottom left). Each pixel in the scattergram (See Fig \ref{Fig5} bottom right) represents a couple of X-intensity and Y-intensity but also a couple of modulus and angle. The origin (null modulus) is choosen as the center of the image. Finally the intensity of each pixel of the scattergram represents the number of times the corresponding couple ($\lbrace modulus,angle\rbrace$ or $\lbrace intensity X, intensity Y\rbrace$) appears in the map. If we sum all the values of the scattergram image, we get the number of pixels used for the inital images (mostly 1024x1024 = 1048576).\\
Filling of the scattergram image can be described by the following equation :
\begin{equation}
\label{Eq3}
	S(i,j) = P \left\lbrace i \leqslant m_x  < i+1 \;\&\; j\leqslant m_y < j+1 \right\rbrace 
\end{equation}
where the $S(i,j)$ are the pixels of the scattergram and the $(m_x,m_y)$ are the components of magnetization normalized by the pixel size. $P$ is the probability of a $(m_x,m_y)$ occurence in the image. The scattergram can be also seen as a representation of the back focal plane of the Lorentz lens, the origin of the scattergram being the transmitted beam. Interested readers can also find information in \cite{Bonnet1995}.

\newpage
\section*{References}
\bibliographystyle{unsrt} 
\bibliography{Biblio}

\begin{thebibliography}{10}

\bibitem{DeGraef2001}
M.~De~Graef.
\newblock {\em Magnetic imaging and its applications to Materials}.
\newblock Academic Press, 2001.

\bibitem{Chapman1984}
J.N. Chapman.
\newblock The investigation of magnetic domain structures in thin foils by
  electron microscopy.
\newblock {\em Journal of Physics D : Applied Physics}, 17:623--647, 1984.

\bibitem{kohler2000}
M.~Kohler, T.~Schweinbock, T.~Schmidt, J.~Zweck, G.~Bayreuther, P.~Fischer,
  G.~Schutz, T.~Eimuller, P.~Guttmann, and G.~Schmahl.
\newblock Imaging of sub-100-nm magnetic domains in atomically stacked
  fe(001)/au(001) multilayers.
\newblock {\em Journal of Applied Physics}, 87(9):6481--6483, 2000.

\bibitem{Hino2004}
T.~Hino, Y.~Murakami, D.~Shindo, S.~Okamoto, O.~Kitakami, and Y.~Shimada.
\newblock Observation of magnetic domain structure in fept (001) films by
  electron holography and lorentz microscopy.
\newblock {\em Journal Of The Japan Institute Of Metals}, 68(5):315--319, May
  2004.

\bibitem{Aitchison2001}
P.R. Aitchison, J.N. Chapman, V.~Gehanno, I.S. Weir, M.R. Scheinfein,
  S.~McVitie, and A.~Marty.
\newblock High resolution measurement and modelling of magnetic domain
  structures in epitaxial fepd (0 0 1) l10 films with perpendicular
  magnetisation.
\newblock {\em Journal of Magnetism and Magnetic Materials}, 223:138--146,
  2001.

\bibitem{Gehanno1997}
V.~Gehanno, Y.~Samson, A.~Marty, B.~Gilles, and A.~Chamberod.
\newblock Magnetic susceptibility and magnetic domain configuration as a
  function of the layer thickness in epitaxial fepd(001) thin films ordered in
  the l1(0) structure.
\newblock {\em Journal of Magnetism and Magnetic Materials}, 172(1-2):26--40,
  August 1997.

\bibitem{Viret2000}
Viret, Samson, Warin, Marty, Ott, Sondergard, Klein, and Fermon.
\newblock Anisotropy of domain wall resistance.
\newblock {\em Phys Rev Lett}, 85(18):3962--3965, Oct 2000.

\bibitem{Chapman1978}
J.~N. Chapman, P.~E. Batson, E.~M. Wadell, and R.~P. Ferrier.
\newblock Direct determination of magnetic domain-wall profiles by differential
  phase-contrast electron microscopy.
\newblock {\em Ultramicroscopy}, 3(2):203--214, 1978.

\bibitem{Daykin1995}
A.C. Daykin and A.K. Petford-Long.
\newblock Quantitative mapping of the magnetic induction distribution using
  foucault images formed in a transmission electron microscope.
\newblock {\em Ultramicroscopy}, 58:365--380, 1995.

\bibitem{AHARONOV1959}
Y.~Aharonov and D.~Bohm.
\newblock Significance of electromagnetic potentials in the quantum theory.
\newblock {\em Physical Review}, 115(3):485--491, 1959.

\bibitem{Teague1983}
M.R. Teague.
\newblock Deterministic phase retrieval : a green's function solution.
\newblock {\em Journal of the Optical Scociety of America}, 73:1434--1441,
  1983.

\bibitem{Paganin1998}
D.~Paganin and K.A. Nugent.
\newblock Noninterferometric phase imaging with partially coherent light.
\newblock {\em Physical review letters}, 80:2586--2589, 1998.

\bibitem{De2001}
M.~De~Graef and Y.~Zhu.
\newblock Quantitative noninterferometric lorentz microscopy.
\newblock {\em Journal of Applied Physics}, 89:7177--7179, 2001.

\bibitem{Volkov2002}
V.V. Volkov, Y.~Zhu, and M.~De~Graef.
\newblock A new symmetrized solution for phase retrieval using the transport of
  intensity equation.
\newblock {\em Micron}, 33:411--416, 2002.

\bibitem{Beleggia2004}
M.~Beleggia, M.A. Schofield, V.V. Volkov, and Y.~Zhu.
\newblock On the transport of intensity technique for phase retrieval.
\newblock {\em Ultramicroscopy}, 102:37--49, 2004.

\bibitem{McVitie2006}
S.~McVitie and M.~Cushley.
\newblock Quantitative fresnel lorentz microscopy and the transport of
  intensity equation.
\newblock {\em Ultramicroscopy}, 106(4-5):423--431, March 2006.

\bibitem{Petersen2007}
Tim~C. Petersen and Vicki~J. Keast.
\newblock Astigmatic intensity equation for electron microscopy based phase
  retrieval.
\newblock {\em Ultramicroscopy}, 107(8):635--643, August 2007.

\bibitem{Koch2008}
Christoph~T. Koch.
\newblock A flux-preserving non-linear inline holography reconstruction
  algorithm for partially coherent electrons.
\newblock {\em Ultramicroscopy}, 108(2):141--150, January 2008.

\bibitem{Volkov2004}
V.~V. Volkov and Y.~Zhu.
\newblock Lorentz phase microscopy of magnetic materials.
\newblock {\em Ultramicroscopy}, 98:271--281, 2004.

\bibitem{Petersen2008}
Tim~C. Petersen, Vicki~J. Keast, and David~M. Paganin.
\newblock Quantitative {TEM-based} phase retrieval of {MgO} nano-cubes using
  the transport of intensity equation.
\newblock {\em Ultramicroscopy}, 108(9):805--815, August 2008.

\bibitem{Dunin-Borkowski2004}
R.~E. Dunin-Borkowski, T.~Kasama, A.~Wei, S.~L. Tripp, M.~J. Hytch, E.~Snoeck,
  R.~J. Harrison, and A.~Putnis.
\newblock Off-axis electron holography of magnetic nanowires and chains, rings,
  and planar arrays of magnetic nanoparticles.
\newblock {\em Microscopy Research And Technique}, 64(5-6):390--402, August
  2004.

\bibitem{Toussaint2002}
J.~C. Toussaint, A.~Marty, N.~Vukadinovic, J.~B. Youssef, and M.~Labrune.
\newblock A new technique for ferromagnetic resonance calculations.
\newblock {\em Computational Materials Science}, 24(1-2):175--180, May 2002.

\bibitem{Malis1988}
T.~Malis, S.~C. Cheng, and R.~F. Egerton.
\newblock Eels log-ratio technique for specimen-thickness measurement in the
  tem.
\newblock {\em J Electron Microsc Tech}, 8(2):193--200, Feb 1988.

\bibitem{Gehanno1997a}
V.~Gehanno, A.~Marty, B.~Gilles, and Y.~Samson.
\newblock Magnetic domains in epitaxial ordered fepd(001) thin films with
  perpendicular magnetic anisotropy.
\newblock {\em Physical Review B}, 55(18):12552--12555, May 1997.

\bibitem{Kruse2006}
P.~Kruse, M.~Schowalter, D.~Lamoen, A.~Rosenauer, and D.~Gerthsen.
\newblock Determination of the mean inner potential in iii-v semiconductors, si
  and ge by density functional theory and electron holography.
\newblock {\em Ultramicroscopy}, 106(2):105 -- 113, 2006.

\bibitem{HUMPHREY1985}
F.~B. Humphrey and J.~C. Wu.
\newblock Vertical bloch line memory.
\newblock {\em Ieee Transactions On Magnetics}, 21(5):1762--1766, 1985.

\bibitem{Thiaville2001}
A.~Thiaville, J.~Miltat, and J.~Ben~Youssef.
\newblock Dynamics of vertical bloch lines in bubble garnets: Experiments and
  theory.
\newblock {\em European Physical Journal B}, 23(1):37--47, September 2001.

\bibitem{Bonnet1995}
N.~Bonnet, M.~Herbin, and P.~Vautrot.
\newblock Extension of the scatterplot approach to multiple images.
\newblock {\em Ultramicroscopy}, 60:349--355, 1995.

\end{thebibliography}

\newpage
\section*{Figures}

\end{document}